\documentclass[aps,pre,preprint,superscriptaddress,showpacs,floatfix]{revtex4}
\usepackage{graphicx}
\begin{document}
\title{Analytical solution of 1D lattice gas model with infinite number of
multiatom interactions}
\author{V. I. Tokar}
\affiliation{
IPCMS-GEMM, UMR 7504 CNRS, 23, rue du Loess,
F-67037 Strasbourg Cedex, France
}
\affiliation{Institute of Magnetism, National Academy of Sciences,
36-b Vernadsky str., 03142 Kiev-142, Ukraine}
\author{H. Dreyss\'e}
\affiliation{
IPCMS-GEMM, UMR 7504 CNRS, 23, rue du Loess,
F-67037 Strasbourg Cedex, France
}
\date{\today}
\begin{abstract}
We consider a 1D lattice gas model in which the atoms interact via an
infinite number of cluster interactions within contiguous atomic chains
plus the next nearest neighbor pairwise interaction. All interactions
are of arbitrary strength. An analytical expression for the size
distribution of atomic chain lengths is obtained in the framework of
the canonical ensemble formalism.  Application of the exact solution to
the problems of self-assembly and self-organization is briefly
discussed.
\end{abstract}
\pacs{68.65.-k, 68.66.La, 81.16.Dn}
\maketitle

\section{Introduction}
In recent years there has been growing interest in studies of monatomic
chains obtained in the processes of heteroepitaxial growth
\cite{chains0,Ti_C,nat_chains} at various substrates.  A major goal of
these studies is to obtain the so-called quantum wires for potential
use in microelectronic applications.  In these experiments one usually
aims at obtaining wires of infinite length for subsequent use in the
studies of the Luttinger liquid \cite{nat_chains}.  In practice,
however, one frequently encounters the monatomic wires of finite
length. These may be useful in more practical applications, such as the
microelectronic circuitry \cite{devices}.  Such finite 1D clusters were
observed, e. g., in Refs.  \cite{Pt997,1D2Dexp}.  Current theoretical
interest is in magnetic properties of finite monatomic clusters
\cite{magnetism} in view of their potential use in magnetic memory
devices. The latter application, however, will require the development
of the techniques of mass production of such objects. In this context
it would be interesting to study the possibility of their self-assembly
and self-organization similar to analogous processes in 2D
heteroepitaxial systems \cite{discovery,dots}.

In the present paper we consider an analytic solution for the cluster
size distribution in the framework of a 1D lattice gas model with an
arbitrary number of cluster interactions within contiguous atomic
chains plus the next nearest neighbor pair interaction. Such model can
be justified in the framework of the Frenkel-Kontorova model of
strained epitaxy \cite{cond-mat1} but also can be useful in other
cases, e.  g., in {\em ab initio} approaches, where the cluster
interactions appear because of the many-body nature of the electron
interactions which cannot be reduced to the pair interatomic
potential.

\section{The model}
Let us  consider a lattice gas model defined by the function $E(l)$
describing the dependence of the energy of 1D atomic clusters (or
chains) on their length $l$. Such a model was already used  for the
description of the self-assembly phenomena in Ref.\ \cite{lannoo}. We
note, however, that in 2D this approach can be only phenomenological
because the cluster size does not characterize a 2D structure
uniquely.  In contrast, in 1D the cluster is defined unambiguously by
its length and the function $E(l)$ is also uniquely defined provided
the clusters do not interact.

To apply standard tools of statistical physics we need to find the
Hamiltonian corresponding to our model. To this end we
consider the following trial expression
\begin{equation} 
\label{Frelax2}
H=\sum_{i;l=2}^\infty V_{l}n_{i}n_{i+1}\dots
n_{{i} + (l-1)},
\end{equation}
where $V_{l}$ are numerical coefficients to be fitted to reproduce the
chain energies $E(l)$ and $n_i$ the occupation numbers taking values 0
or 1 depending on whether site $i$ is empty or is occupied by an atom.
Assuming we have already fitted $V_{l-1}, V_{l-2},\dots, V_{2}$ to the
energies $E(l-1)\equiv E_{l-1}, E_{l-2},\dots, E_{2}$ let us consider
the chain of length $l$. From Eq.\ (\ref{Frelax2}) one finds
\begin{equation} 
\label{l+1}
E(l)=E(l-1)+V_{l}+V_{l-1}+V_{l-2}+\dots+V_{2},
\end{equation}
where the $V$ terms on the right hand side (r. h. s.) account for the
interactions of the newly added atom with the rest.  Similarly
\begin{equation} 
\label{l}
E(l-1)=E(l-2)+V_{l-1}+V_{l-2}+\dots+V_{2}.
\end{equation} 
Subtracting Eq.\ (\ref{l}) from Eq.\ (\ref{l+1})
we get
\begin{equation}
E(l)-E(l-1)=E(l-1)+V_{l}-E(l-2)
\end{equation}
from where it follows that the Hamiltonian of our model is
Eq.\ (\ref{Frelax2}) with the coefficients given by the recursion
relation
\begin{equation}
\label{V_l}
V_l=E_l-2E_{l-1}+E_{l-2}
\end{equation}
initialized by $E_0=E_1=0$. For the system to be well defined in the
thermodynamic limit the chain energy $E_l$ cannot grow quicker then
linearly when $l\to\infty$. This means that the the cluster
interactions $V_l$ tend to zero at large $l$ because according to
Eq.\ (\ref{V_l}) they are equal to the discrete second derivative of
$E_l$ with respect to $l$.

In the above model atoms interact only when they belong to the same
contiguous chain, so that separate chains are not coupled to each
other. To make the model more realistic we allow for the interchain
coupling by adding to the Hamiltonian (\ref{Frelax2}) the
next nearest neighbor (NNN) pair interaction term
\begin{equation} 
\label{ H}
H=V_{NNN}\sum_in_in_{i+2}+\sum_{i;l=2}^\infty V_{l}n_{i}n_{i+1}\dots
n_{{i} + (l-1)}.
\end{equation} 

\section{The canonical ensemble solution}
We consider a 1D lattice with a fixed number of atoms $N$ which can
occupy $I>N$ lattice sites. Our major goal in this paper is to find the
equilibrium distribution of the atomic chain sizes at finite
temperature $T$ in the model defined by Hamiltonian (\ref{ H}).  But
first we will consider the above problem using a simpler Hamiltonian
(\ref{Frelax2}), i. e., by neglecting the NNN interaction which will be
accounted for later. We will seek the canonical ensemble solution of
our problem by generalizing the approach of Ref.\ \cite{vavro}.  In
this approach one first have to compute the energy of the system in
terms of cluster variables $m_l$---the number of clusters of length $l$
and their energy $E(l)$.  Because the $l\rightarrow\infty$ limit is not
completely trivial, we calculate the configurational energy by assuming
that all ${V}_l$ with $l>L$ are equal to zero. In this case the total
energy can be calculated as
\begin{equation} 
\label{ E}
{\cal E} = \sum_{l=1}^{L-2}E_lm_l + E_{L}^{\prime}N_> 
-m_>\sum_{l=2}^{L}(l-1){V}_l,
\end{equation} 
where $E_{L}^{\prime} =\sum_{l=2}^{L}{V}_l$ is the first (discrete)
derivative of $E_l$, $m_>$ is the total number of clusters of length
exceeding $L-2$ which below we will call the big clusters, and
$N_>=\left(N- \sum_{l=1}^{L-2}lm_l \right)$ is the number of atoms
contained in the big clusters. The last two terms on the r. h. s. of
Eq.~(\ref{ E}) can be verified by first checking their correctness for
$m_>=1$ and then noting that cutting a big cluster into two ones
amounts to cutting $l-1$ couplings of type ${V}_l$. To complete the
calculation of the total free energy $F_{{tot}} = E_{{tot}}  - TS$ it
remains to compute the entropy $S$. The number of atomic configurations
can be expressed through the cluster variables as
\begin{equation} 
\label{ Omega1}
\Omega_{ at}=\frac{m_{ tot}!}{m_1!m_2!\ldots m_>!}
\left(\begin{array}{c} 
N_{>}-(L-2)m_{>}-1 \\ 
m_{>}-1 
\end{array}\right)
\end{equation} 
In this equality the first multiplier on the r. h. s.  counts all
inequivalent permutations between the clusters with all big clusters
being considered as equivalent while the second factor accounts for the
number of ways to divide $N_>$ atoms into $m_>$ big clusters including
their permutations.  We find this factor by observing that the division
of $N_>$ atoms into clusters of sizes exceeding $L-2$ is equivalent to
dividing $n=N_>-(L-2)m_>$ atoms into $k=m_>$ clusters. The latter
quantity was calculated in Ref.  \cite{vavro} to be equal to the
binomial coefficient
\[C_{k-1}^{n-1}=\left(  \begin{array}{c}
          n-1\\
              k-1
        \end{array} \right)\] 
provided $n\geq k\geq1$. Otherwise it is equal to zero.  The meaning of
this formula is simple. In a contiguous chain of $n$ atoms there is
$n-1$ places to cut the chain into pieces. $k$ pieces can be obtained
with $k-1$ cuts. Hence, the number of the possible cuts of $n$ atoms
into $k$ pieces is equal exactly to the above combinatorial
coefficient.  To complete the calculation of the total number of
configurations we have to multiply $\Omega_{ at}$ by the corresponding
factor responsible for the configurations of vacant sites~\cite{vavro}.
But before doing this we repair Eq.~(\ref{ E}) by accounting for the
omitted NNN pair interaction. To this end we note that as long as the
contiguous chains are concerned, $V_{{NNN}}$ interaction amounts simply
to renormalization of $V_3$. Additionally this term introduces the
interchain coupling between the chains separated by a single vacant
site (a vacancy). Therefore, the term $V_{{NNN}}k_1$, where $k_1$ the
number of vacancies, should be added to the r. h. s. of Eq.~(\ref{ E})
(here and above we do not pay attention to the boundary
conditions~\cite{vavro} because in this paper we are interested only in
the thermodynamic limit). So when counting the configurations of empty
sites we should separate the configurations with different numbers of
vacancies. This is achieved by using Eq.~(\ref{ Omega1}) with the total
number of empty sites $I-N$ instead of $N$, the big clusters in this
case should exceed $L-2 = 1$. Denoting the total number of clusters as
$k$ from Eq.~(\ref{ E}) we get
\begin{equation} 
\label{ Omega2}
\Omega_{ vac}=\frac{k!}{k_1!(k-k_1)!}
\left(  \begin{array}{c}
          I-L-k-1\\
              k-k_1-1
        \end{array} \right).
\end{equation} 
Our final expression for $S$ is obtained from Eqs.~(\ref{ Omega1}) and
(\ref{ Omega2}) with the use of the Stirling formula as
\begin{eqnarray} 
\label{ S}
&&\frac{S}{k_BI} = 2c  \ln c  -\sum_{l=1}^{L-2}c_l\ln c_l -v\ln v 
- 2(c-v)\ln (c-v)\nonumber\\
&& +(1-\theta-c)\ln(1-\theta-c)-(1-\theta-2c+v)\ln(1-\theta-2c+v)\\
&&-2c_>\ln c_>+[\theta_>-(L-2)c_>]\ln[\theta_>-(L-2)c_>]-[\theta_>-(L-1)c_>]
\ln[\theta_>-(L-1)c_>],\nonumber
\end{eqnarray} 
where $\theta=N/I$ is the total coverage, $c_l=m_l/I$ is the
concentration of clusters of length $l$, $c=\sum_lc_l$ the total
cluster concentration, $v=k_1/I$ the vacancy concentration, $\theta_> =
\theta-\sum_{l=1}^{L-1}lc_l=N_>/I$. This expression for reduced entropy
together with the reduced energy obtained from Eq.~(\ref{ E}) are
sufficient to obtain the reduced free energy as a function of the
unknown quantities $\{c_l\}$, $c_>,$ and $v$. Minimizing it with
respect to these variables we arrive in the limit $L\rightarrow\infty$
at the following set of equations
\begin{eqnarray} 
\label{ c_l}
&&c_l = \frac{c^2(1-\theta-2c+v)^2}{(c-v)^2(1-\theta-c)}
\exp\left({\frac{\mu l-E_l}{k_BT}}\right)\\
\label{ c_l2}
&&(c-v)^2 = v(1-\theta-2c+v)\exp\left(\frac{V_{NNN}}{k_BT}\right).
\end{eqnarray} 
To find the above limit it is necessary to know the asymptotic behavior
of $c_l$ at large $l$.  The limit was taken by assuming the exponential
behavior $c_l\propto\exp(-\lambda l)$ which is consistent with the
resulting equation~(\ref{ c_l}).  We note that Eq.~(\ref{ c_l2}) is of
the second order in $v$, so one can exclude this variable from the
first equation to obtain a closed equation for $\{c_l\}$. From the
above set, however, it is easier to see that if the NNN interaction is
negative, than $v\rightarrow c$ as $T\rightarrow0$, i.~e., the vacancy
concentration becomes equal to the concentration of clusters which
means that the system becomes ordered: the clusters self-organize into
chains separated by monovacancies. This is further confirmed by the
fact that the entropy Eq.~(\ref{ S}) tends to zero at zero temperature
provided $V_{NNN} < 0$ and the clusters are size calibrated. Indeed, in
the limit $L\to\infty$ only the first two lines of Eq.\ (\ref{ S})
survive
\begin{widetext} 
\begin{eqnarray} 
\label{ s}
{\frac{s}{k_B}}&=& 2c  \ln c  -\sum_{l=1}^{L-2}c_l\ln c_l -v\ln v -
2(c-v)\ln (c-v)\nonumber\\ &&+(1-\theta-c)\ln(1-\theta-c)
-(1-\theta-2c+v)\ln(1-\theta-2c+v),
\end{eqnarray} 
\end{widetext} 
where $s=S/I$ is the entropy per site.

In connection with this expression it is worth noting that Eqs. (\ref{
c_l}) and (\ref{ c_l2}) can formally be obtained in the thermodynamic
limit from the variation with respect to $c_l$ of the
expression for the free energy density 
\[
f=\sum_{l=1}^\infty c_lE_l -k_BTs-\mu(\sum_{l=1}^\infty lc_l-\theta),
\]
where $\mu$ is the Lagrange multiplier. This derivation, however, may
cause doubts in the case of clusters of sizes $l=O(N)$ when the terms
$lc_l=m_lO(N)/N$ in the above equation acquire discrete values and so
are not suitable for the variational treatment. The derivation
presented above is more rigorous and besides can be modified to be
applicable to finite systems \cite{vavro,girifalco}.

At $T=0$ and $c=v$ the entropy density (\ref{ s}) takes the form
\begin{equation} 
{\frac{s}{k_B}}= c  \ln c  -\sum_{l=1}^{L-2}c_l\ln c_l.
\end{equation} 
Thus, if the clusters at $T=0$ are size calibrated, i. e., if for some
value of $l=l_0$ $c_{l_0} = c$ while $c_l=0$ for all other $l$, then
from the above equation it follows that $s=0$.  For $V_{NNN} \geq 0$
the entropy is positive even at $T=0$ meaning disordered state.  The
issue of size calibration will be considered in more detail in section
\ref{sizecalibration}.

\section{The Ising model}
To check Eqs.\ (\ref{ c_l}) and (\ref{ c_l2}) we will apply them to the
known exactly solvable problem---the 1D Ising model.  As is known, it
is equivalent to the lattice gas model with pair interatomic
interaction. This can be easily shown (see \cite{ducastelle} Ch. II) by
substituting $n_i=(1-\sigma_i)/2$ ($\sigma_i=\pm1$---the Ising spin
variable) into the LGM Hamiltonian
\begin{eqnarray}
\label{ HIM}
H&=&\frac{1}{2}\sum_{ij}V_{ij}n_in_j=\frac{1}{8}\sum_{ij}V_{ij}
(1-\sigma_i)(1-\sigma_j)\nonumber\\
&=& \frac{1}{2}\sum_{ij}(V_{ij}/4)\sigma_i\sigma_j
+\sum_j(V_{ij}/4)(1-2\sigma_j)
\end{eqnarray}
Thus, the Ising model with NN interaction can be solved in the
canonical ensemble formalism with the use of our formulas.  In this
section we will obtain the appropriate formulas in closed analytical
form. Also, we will compare our results obtained in the canonical
ensemble with more familiar results obtained in the grand canonical
formalism via the transfer matrix method. The latter solution can be
found  e.\ g., in Ch. V of Ref.\ \cite{ducastelle}, where the 1D Ising
model Hamiltonian was considered which in our notation can be written
as (cf. Eq.\ (\ref{ HIM})
\begin{equation} 
\label{ HIM2}
H_{IM}=(V_{NN}/4)\sum_i\sigma_i\sigma_{i+1}
-\frac{1}{2}h\sum_i(\sigma_i+\sigma_{i+1})
\end{equation}
was considered. The reduced free energy calculated in the grand
canonical ensemble was found to be
\begin{equation} 
\label{ F2}
\frac{F_{GC}}{Nk_BT}=-\ln\left(e^{-K/4}\cosh\bar{h}
+\sqrt{e^{-K/2}\sinh^2\bar{h}+e^{K/2}}\right),
\end{equation} 
where $K=V_{NN}/k_BT$ and $\bar{h}=h/k_BT$ is the external field
(divided by $k_BT$) which fixes the magnetization $M$. The latter is
connected to the coverage $\theta$ by the relation
\begin{equation} 
\label{ thetaM}
\theta = (1-M)/2
\end{equation}
and can be calculated as
\begin{equation} 
\label{ M}
M=\partial F_{GC}/\partial \bar{h}.
\end{equation} 
To obtain the canonical ensemble free energy we have to perform the
Legendre transform (see Eq.\ (\ref{ HIM2}))
\begin{equation} 
\frac{F_C}{Nk_BT}=F_{GC}+\bar{h}\frac{\partial F_{GC}}{\partial
\bar{h}},
\end{equation} 
where the partial derivative is calculated from Eq.\ (\ref{ M}) as
\begin{equation} 
\label{ M2}
M = \frac{e^{-K/4}\sinh\bar{h}(1+e^{-K/4}\cosh\bar{h}/
\sqrt{e^{-K/2}\sinh^2\bar{h}+e^{K/2}})}{e^{-K/4}\cosh\bar{h}
+\sqrt{e^{-K/2}\sinh^2\bar{h}+e^{K/2}}}.
\end{equation} 
Finally, to obtain the LGM free energy $F$ in the canonical ensemble
formalism we have to add the statistical average of the last term on 
the r. h. s. of Eq.\ (\ref{ HIM}):
\begin{equation} 
\label{ F}
\frac{F}{Nk_BT}=\frac{F_C}{Nk_BT}+(K/4)(1-2M).
\end{equation} 

In our formalism the free energy $F$ for the NN LGM can be calculated as
follows. According to Eq.\ (\ref{ c_l2}), in the case $V_{NNN}=0$
\begin{equation}
\label{ v}
v=c^2/(1-\theta).
\end{equation}
Substituting this into Eq.\ (\ref{ c_l}) we get in the case of the
Ising model
\begin{equation} 
\label{ c_l1}
c_l = (1-\theta-c)\exp[\bar{\mu}+(\bar{\mu}-K)(l-1)],
\end{equation} 
where $\bar{\mu}=\mu/k_BT$.
From Eq.\ (\ref{ c_l1}) we easily find
\begin{equation}
\label{ c}
\left\{
\begin{array}{rcl}
\displaystyle{c}&=&\displaystyle{\sum_{l=1}^{\infty} c_l
=}\displaystyle{(1-\theta-c)e^{\bar{\mu}}/(1-e^{\bar{\mu}-K})}
\nonumber\\
\theta&=&\displaystyle{\sum_{l=1}^\infty 
lc_l=}\displaystyle{c+c^2e^{-K}/(1-e^{\bar{\mu}-K}).}
\end{array}\right.
\end{equation} 
These two equations are sufficient to express the unknown quantities
$c$ and $\bar{\mu}$ through the coverage $\theta$ and the interaction
parameter $K$---the independent variables in the canonical ensemble.
From Eq.\ (\ref{ c}) we have after some algebra 
\[
\left\{
\begin{array}{rcl}
\bar{\mu}&=&\displaystyle{-\ln\left[\frac{1}{2\theta}
-x+\sqrt{\left(\frac{1}{2\theta}
-x\right)^2+x(1-x)}\right]}\\
c&=&\displaystyle{\frac{1}{2x}\left[1-\sqrt{1-4\theta(1-\theta)x}\right]},
\end{array}
\right.
\]
where $x=(1-e^{-K})$. These expressions can be used to obtain analytic
expressions for all quantities of interest.

In the canonical ensemble case the free energy is
\begin{equation} 
\label{ F1}
F=k_BT\left[\bar{\mu}\theta-(1-\theta)\ln\left(1+\frac{e^{\bar{\mu}}}{1-
e^{\bar{\mu}-K}}\right)\right].
\end{equation}
As we see, it is quite different from Eq.\ (\ref{ F}). We were unable
to compare the two solutions analytically and used the numerical
procedure. By fixing $\bar{h}$ and $K$ we computed $M$ with the use of
Eq.\ (\ref{ M2}) and $F/Nk_BT$ from Eq.\ (\ref{ F}). Then $\theta$ was
calculated according to Eq.\ (\ref{ thetaM}) and $F/Nk_BT$ was computed
from Eq.\ (\ref{ F1}). Both values coincided within 13-15 significant
digits depending on how singular the values of $F/Nk_BT$ are at the
chosen values of $\bar{h}$ and $K$.

Thus we have solved the 1D NN LGM in the canonical ensemble approach
and expressed all quantities of interest in terms of the atomic density
(or coverage) $\theta$ which is more natural and easier to measure
quantity then the chemical potential variable of the grand
canonical formalism.
\section{The size calibration}
\label{sizecalibration}
According to Ref.\ \cite{lannoo}, the system exhibits the size
calibration of the atomic clusters if the energy density per atom
$E(l)/l$ has a minimum.  The size calibration is easiest to see in the
case $V_{NNN} = 0$ when the distribution Eq.~(\ref{ c_l}) takes the
form \[ c_l = (1-\theta-c)\exp\beta(\mu l - E_l).\] Except for the
excluded volume prefactor this expression coincides with the formula
proposed in Ref.\ \cite{lannoo} where a detailed discussion of the size
calibration issue is given and which fully applies to our case provided
the pair attraction $V_{NN}$ is not too strong. In
Ref.\ \cite{cond-mat1} we have shown that in the framework of the
Frenkel-Kontorova model the effective chain energy (in fact, the free
energy---see Ref.\ \cite{cond-mat1} for details) has the form
\begin{equation}
\label{Frelax}
E_l= V_{NN}(l-1)-\sum_{\text{chains}}(W_l-TS_l)
\end{equation}
where the last two terms represent the relaxation free energy.  The
length dependence of the relaxation energy $W_l$ is governed by the
dimensionless parameter $\alpha=k_s/k_p$, where the spring constants
$k_s$ and $k_p$ are the second derivatives of the potentials which bind
the atom to the substrate ($k_s$) and of the pair interatomic
interaction ($k_p$), respectively.  In the illustrative calculations
below the energy unit was chosen to be equal to the energy of the
unrelaxed misfit strain $k_pf^2$, where the interatomic spring constant
$k_p$ was defined earlier and $f$ is the misfit as defined in
Ref.\ \cite{cond-mat1}.  The relaxation entropy $S_l$ is practically
linear in $l$ and so according to Eq.\ (\ref{V_l}) essentially
contributes only into the NN interaction.

\begin{figure}
\includegraphics[viewport = 195 520 415 690, scale = 0.9]{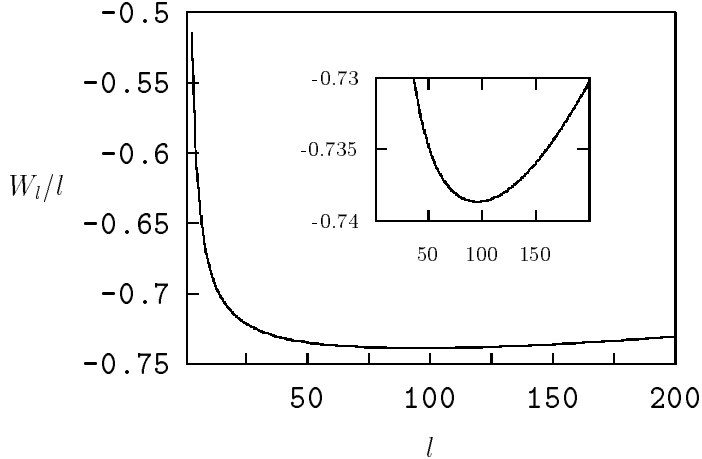} 
\caption{The length dependence of the reduced chain energy with the 
relaxation energy corresponding to $\alpha=10^{-5}$ and $V_{NN}=-0.25$
in units of $k_pf^2$ (see the text). 
The inset shows the location of the minimum.}
\label{fig1}
\end{figure}

\begin{figure}
\includegraphics[viewport = 195 520 415 740, scale = 0.7]{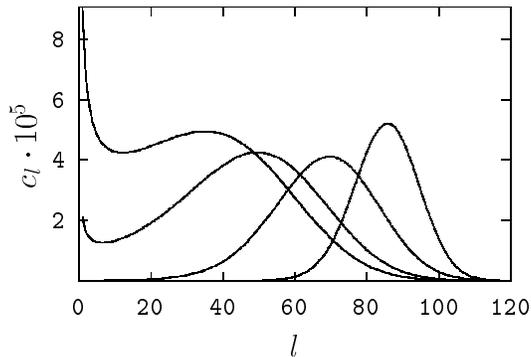} 
\caption{Equilibrium distribution of atomic chain lengths at the
coverage $\theta = 0.1$ corresponding to (from left to right) 
$1/k_BT=$20, 22, 30, and 60 (in units of $k_pf^2$). The pair interaction
$V_{NN}=-0.25$ and $\alpha=10^{-5}$ 
(for explanation of notation see the text).}
\label{fig2}
\end{figure}

\begin{figure}
\includegraphics[viewport = 195 520 415 740, scale = 0.7]{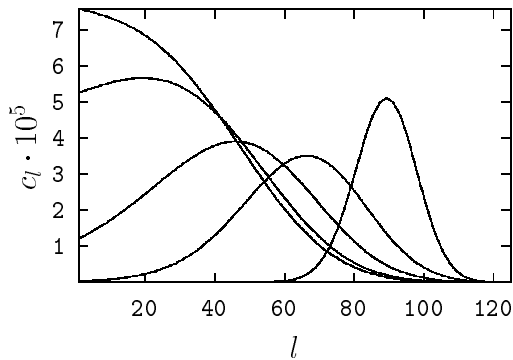} 
\caption{The same as in Fig.\ \ref{fig2} except that the entropic part
of $E_l$ is set to zero and the temperatures are (from left to right)
$1/k_BT=$12.5, 13, 15, 20, and 60
}
\label{fig3}
\end{figure}

In our calculations we chosen $\alpha=10^{-5}$ to be quite small in
order to visualize all qualitative details of the self-assembly
behavior. Physically this would correspond to very weak binding to the
substrate. In the above energy units $V_{NN}$ was chosen to be equal to
-0.25 so that there was a minimum in $E(l)/l$ at $l\approx100$ (see
Fig.\ref{fig1}). The calculation of the size distribution for this case
is shown in Fig.\ \ref{fig2}. It qualitatively agrees with the Monte
Carlo simulations of Ref.\ \cite{shchukin} (cf. Fig.\ 2 of that
reference).  The most notable feature of the above calculations is the
change of the position of the maximum of the cluster size distribution
with lowering temperature---the feature not clearly understood from the
theory of Ref. \cite{lannoo}.
Although the entropic contribution into $E_l$ is rather small due to
the small temperatures considered, the shallowness of the minimum in
$E(l)/l$ makes it possible that the above shift of the maximum is due
to the shift in the position of the minimum in $E_l=W_l-TS_l$ with
lowering temperature. To check this possibility we repeated
calculations with $E_l$ replaced by temperature independent part $W_l$.
The results of this calculation is shown in Fig.\ \ref{fig3}. The only
qualitative difference with Fig.\ \ref{fig2} is that the bimodal
distribution is never observed in the case of $T$-independent case.
Presumably, this is due to the 1D character of our model, because in 2D
case where also $T$-independent interactions were used we found a
strongly bimodal distribution \cite{molphys}.  
\section{Conclusion}
In this paper we considered an analytical solution of a 1D lattice gas
model with an arbitrary number of cluster interatomic interactions
within contiguous atomic chains. We illustrated the model by applying
it to the description of the equilibrium distribution of atomic
clusters in strained epitaxy in the thermodynamic limit.  However, the
model can be used also in other cases when the multiatom interactions
are of the above type, as well as applied to finite systems, like those
discussed in Refs.\ \cite{vavro,girifalco}. Our analytical solution
confirms the observation made in Ref.\ \cite{shchukin} in the framework
of a phenomenological approach that the maximum in the {\em
equilibrium} distribution of the cluster sizes moves to higher values
with lowering temperature. This can lead to an interesting kinetic
phenomenon consisting in the saturated ripening when the system is
quenched to lower temperature.
\begin{acknowledgments}
One of us (VT) expresses his gratitude to University Louis Pasteur de
Strasbourg and IPCMS for their hospitality.
\end{acknowledgments}

\end{document}